\documentclass[pra]{revtex4-1}
\usepackage{graphicx}
\begin{document}
\title{Pattern Formation in a Simple Model of Photochemical Reaction}
\author{Hidetsugu Sakaguchi, Daishiro Kijima,  Shunsuke Chatani$^1$, 
and Toshiaki Hattori$^1$}
\affiliation{Department of Applied Science for Electronics and Materials,
Interdisciplinary Graduate School of Engineering Sciences, Kyushu
University, Kasuga, Fukuoka 816-8580, Japan\\
$^1$ Mitsubishi Rayon Co.,Ltd., Tsurumi, Yokohama, Kanagawa 230-0053, Japan}
\begin{abstract}
Stripe and columnar patterns were experimentally found in photopolymer films during irradiation of collimated UV light. We propose a mathematical model of photopolymerization, perform numerical simulation, and find a spatially periodic columnar structure.
\end{abstract}
\maketitle
Recently, the self-organization of microstructures has been intensively studied in physics, material science, and chemistry.~\cite{rf:1,rf:2} 
Self-organization and pattern formation have also been studied in polymer science. There are various mechanisms of self-organization in polymer science.  Spatially periodic Turing  patterns were found in reaction-diffusion systems in gels.~\cite{rf:3,rf:4}  Honeycomb structures were formed when an amphiphilic polymer solution was cast on solid surfaces, which was due to the Marangoni convection.~\cite{rf:5} Various patterns were created on the surface of polymer gels  during  the volume phase transition owing to mechanical instability.\cite{rf:6} Various mesoscopic patterns were found in block copolymers by micro phase separation.~\cite{rf:7}

In this study, we consider self-organization in photopolymerization.  In photopolymerization, UV light is illuminated onto a monomer solvent including photoinitiators. 
The initiators are excited by UV light and radicals are created. 
A monomer with a radical is added to a polymer and the polymerization chain grows. 
Photopolymerization stops when radicals become inactive or markedly decrease in number. There are many applications of photopolymers, i.e., a UV curing resin and a photoresist for lithography.  Various types of pattern formation were reported in photo-crosslinking and photopolymerization reactions. For example, Nakanishi et al. found a hexagonal phase during the synthesis of interpenetrating polymer networks in a polymer mixture. They concluded that pattern formation occurs by the cooperation between  phase separation and elastic repulsion.~\cite{rf:8} 

On the other hand, Honda et al. found stripe patterns in photopolymer films during the irradiation of tube UV light.~\cite{rf:9} 
Uematsu et al. found amorphous columnar patterns using point light.~\cite{rf:10} In their experiment, a monomer mixture of methacrylate monomer and  photoinitiator was injected into the space between two coverglasses, and  UV light was irradiated onto the setup. The thickness of the space was controlled. No patterns appeared in a thin sample of 0.45 mm, but patterns appeared in a thick sample of 1.5 mm thickness. In the thick sample, a pattern was generated below a depth of about 200 $\mu$m from the irradiated surface. The wavelength of the stripe pattern is about 10 $\mu$m.  Chatani and Hattori performed a similar type of experiment using different monomer mixtures and confirmed columnar patterns. They found that no pattern appeared in a thin sample and  a clear pattern in a medium-thickness sample; however the pattern became random in a thick sample.~\cite{rf:11}  Pattern formation seems to occur during photopolymerization but its detailed mechanism is not yet understood. Honda et al. suggested a diffraction effect and Uematsu et al. proposed a type of lens effect as a mechanism of pattern formation.  That is, if the intensity of  UV light is inhomogeneous  as a result of some perturbations, photopolymerization proceeds inhomogeneously and the spatial inhomogeneity of the refraction index increases, which makes the inhomogeneity of the UV light intensity increase further by the diffraction and (or) refraction effect.  This type of instability in the photochemical reaction has not yet been studied mathematically.    

We numerically study a mathematical model of photochemical polymerization. We assume  radical polymerization induced by photoinitiators. Firstly, radicals, R, are generated by the decomposition of initiators, I, by UV radiation. The addition of a radical to a monomer, M, produces a propagating radical, Q.  The chain propagation occurs by the addition of a monomer to the propagating radical. The reaction of two propagating radicals terminates the chain propagation and yields a polymer, P. The reactions are expressed as 
\begin{eqnarray}
{\rm I}&\rightarrow & 2{\rm R},\nonumber\\
{\rm R}+{\rm M}&\rightarrow & {\rm Q},\nonumber\\
{\rm Q}+{\rm M}&\rightarrow & {\rm Q},\nonumber\\
2{\rm Q}\rightarrow &P
\end{eqnarray}
We assume that the rate of the production of radicals from initiators is proportional to the light intensity $|\phi|^2$. 
The reaction rate equations are expressed as
\begin{eqnarray}
\frac{d I}{d t}&=&-k_1|\phi|^2I,\nonumber\\
\frac{dR}{dt}&=&2k_1|\phi|^2I-k_2R M,\nonumber\\
\frac{dM}{dt}&=&-k_2R M-k_3MQ,\nonumber\\
\frac{dQ}{dt}&=&k_2R M-2k_4Q^2,\nonumber\\
\frac{dP}{dt}&=&k_4Q^2,
\end{eqnarray}
where $I,R,\cdots,P$ are the concentratios of I, R,$\cdots$, and P, and $k_1,\cdots,k_4$ are reaction rate constants. 
The concentration of I decreases monotonically and becomes zero finally. 
The polymer concentration $P$ increases monotonically and becomes constant 
by the termination of the reactions. The stationary state depends on the initial conditions, which is rather different from the simple model studied in  previous sections. 
This is because the polymerization reaction expressed by eq.~(1) is irreversible.  

We have performed numerical simulation of eq.~(1). Figure 1(a) shows  $P$ in the stationary state for $k_1=0.005$ and $k_2=k_3=k_4=1$ as a function of the light intensity $|\phi|^2$. The initial conditions of $I$ are set to be $I(0)=1$, 0.3, and 0.1. The other initial conditions are $M(0)=1$, $R(0)=0$, $Q(0)=0$, and $P(0)=0$. The stationary value of $P$ increases rapidly near $|\phi|^2=0$ with the light intensity for the initial conditions, because the production of radicals  from photoinitiators is proportional to the light intensity.

To study pattern formation, we investigate the following coupled reaction-diffusion equations:
\begin{eqnarray}
\frac{\partial I}{\partial t}&=&-k_1|\phi|^2I+D_I\nabla^2I,\nonumber\\
\frac{\partial R}{\partial t}&=&2k_1|\phi|^2I-k_2R M+D_R\nabla^2R,\nonumber\\
\frac{\partial M}{\partial t}&=&-k_2R M-k_3MQ+D_M\nabla^2Q,\nonumber\\
\frac{\partial Q}{\partial t}&=&k_2R M-2k_4Q^2,\nonumber\\
\frac{\partial P}{\partial t}&=&k_4Q^2,
\end{eqnarray}
where $D_I$, $D_R$, and $D_M$ are diffusion constants for I, R, and M, respectively, and the diffusion of macromolecules Q and P is neglected.  
These reaction-diffusion equations are coupled with the Schr\"odinger-type equation for the light propagation:
\begin{equation}
i\frac{\partial \phi}{\partial z}=-\frac{1}{2}\nabla_2^2\phi-\alpha P\phi.
\end{equation}
Here, we have assumed that the refractive index is proportional to the polymer concentration $P$. 

We have performed numerical simulation in a two dimensional system of size $L_x\times L_z$. The parameters are set to be $k_1=0.005$, $k_2=k_3=k_4=1$, $D_I=D_R=D_M=1$, and $\alpha=200$.  
The input light intensity is assumed to be uniform $\phi(x,z)=1$ at $z=0$. 
 A stationary pattern of $P$ is obtained from the initial condition of $I(x,z)=1-0.1\{1+\cos(24\pi x/L_x)\},M(x,z)=1$, and $R(x,z)=Q(x,z)=P(x,z)=0$ for $L_x\times L_z=80\times 15$. A spatially periodic pattern with $k=24\pi/L_x$ is created from these initial conditions. We have calculated the amplitude $W$ of  spatially periodic structures between $0<z<z_0=5$ by $W=[\int_0^{z_0}\int_0^{L_x} P^2dxdz/(z_0L_x)-\{\int_0^{z_0}\int_0^{L_x} Pdxdz)/(z_0L_x)\}^2]^{1/2}$ for the initial conditions $I(x,z)=1-0.002\{1+\cos(k x)\},M(x,z)=1,R(x,z)=Q(x,z)=P(x,z)=0$ with various initial wavenumbers $k$. Figure 1(b) shows the dependence of $W$ on $k$. The amplitude $W$ takes its maximum near $k=22\pi/L_x$.  This implies that a pattern with a finite wavenumber near $k=22\pi/L_x$ grows dominantly.  
 
\begin{figure}[t]
\begin{center}
\includegraphics[height=4cm]{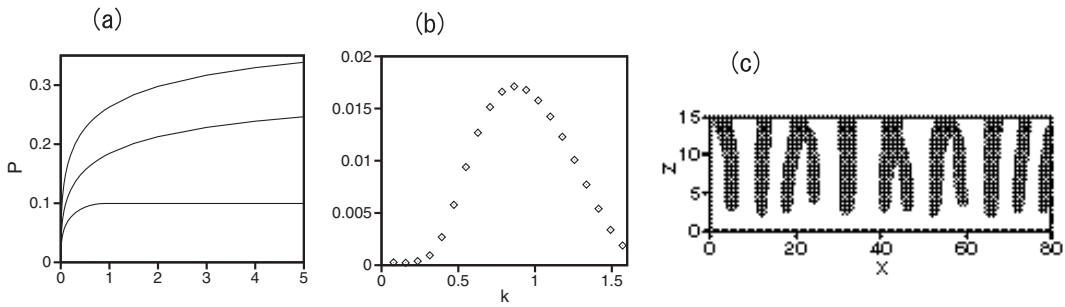}
\end{center}
\caption{(a) Polymer concentration $P$ as function of $|\phi|^2$  in eq.~(1) for the three initial conditions of $I(0)=0.1,0.3$, and 1. The other initial conditions are $M(0)=1$ and $R(0)=Q(0)=P(0)=0$. The parameter values are $k_1=0.005$, $k_2=k_3=k_4=1$, and $D_I=D_R=D_M=1$. (b) Amplitude $W$ of spatially periodic pattern  obtained from the initial condition  $I(x,z)=0.998-0.002\cos(kx)$ with various wavenumbers $k$.
(c) Pattern of polymerization. In the shaded region, $P>0.2625$. The initial condition of $I$ is random between $0.9<I<1$. }
\label{f1}
\end{figure}

We have performed numerical simulation from a random initial condition. That is, $I(x,z)$ is assumed to be a random number between 0.9 and 1, and the other initial condition is the same as before: $M(x,z)=1,R(x,z)=Q(x,z)=P(x,z)=0$. 
The input light is uniform in this numerical simulation, that is, $\phi_0(x)=1$ at $z=0$. Figure 1(c) shows a stationary pattern of $P$ for $L_x\times L_z=80\times 15$. A spatially periodic columnar structure of about 12 columns appears, although some randomness is overlapped. The wavenumber is therefore  $k\sim 24\pi/L_x$. The spatially periodic columnar structure is clearly seen only for $z>2$. Even if the system size is twice enlarged to $L_x\times L_z=160\times 15$, we have obtained a similar pattern of 24 columns.  We think that the fact that a pattern with a finite wavenumber with $k\sim 24\pi/L_x$ is self-organized from a random initial condition is closely related to the numerical result shown in Fig.~1(b) that a pattern with a finite wavenumber tends to grow fast. 

\begin{figure}[t]
\begin{center}
\includegraphics[height=4.cm]{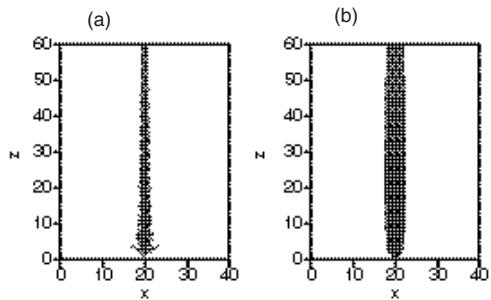}
\end{center}
\caption{(a) Pattern of light intensity in thick system of $L_x\times L_z=40\times 60$. In the shaded region, $|\phi|^2>0.2$. The initial condition of $I$ is 1, and the input light is $\phi_0(x)=0.3/{\rm cosh}\{(x-L_x/2)/6\}$. The parameter value of $\alpha$ is 80. (b) Pattern of polymer. $P>0.6$ in the shaded region. 
}
\label{f2}
\end{figure}

We have performed another type of simulation for the self-organization of a light waveguide in this thicker system. In this simulation, the input light is not uniform, but localized in space. We have assumed that the input light is localized as $\phi_0(x)=0.3/{\rm cosh}\{(x-L_x/2)/6\}$. Figures 2(a) and 2(b) respectively show stationary patterns of $|\phi|^2$ and $P$, where $|\phi|^2>0.2$ and $P>0.6$ are satisfied in the shaded regions.  The parameter $\alpha$ is 80 in this numerical simulation. The beam structure and  strongly polymerized region are localized near $x=L_x/2$ and rather uniform in the $z$-direction. This localized pattern corresponds to a soliton state. 

Similar type of self-organized light waveguide was experimentally studied in the past by Kagami et al., who found  that a light waveguide self-organizes by photopolymerization with beam illumination.~\cite{rf:12} We think that their self-organized light waveguide is closely related to our simulation result.

In summary, we have proposed a  model of  photopolymerization. 
The irreversibility of photopolymerization is incorporated in this model. The reaction rate of the decomposition of initiators is assumed to be proportional to the light intensity, and the complicated reaction-diffusion equations are coupled with  the Schr\"odinger-type equation for light propagation. 
The complicated model equations are difficult to analyze theoretically, but we have performed numerical simulation. A spatially periodic columnar pattern is obtained from a slightly random initial condition of $I(x,z)$, which is qualitatively similar to the pattern formation in the simple model.


\begin{thebibliography}{99}
\bibitem{rf:1} M.~C.~Cross and P.~C.~Hohenberg: Rev. Mod. Phys. {\bf 65} (1993) 851.
\bibitem{rf:2} M.~C.~Cross and H.~Greenside: {\it Pattern Formation and Dynamics in Nonequilibrium Systems} (Cambridge University Press, 2009).
\bibitem{rf:3} V.~Castets, E.~Dulos, J.~Boissonade, and P.~De Kepper:  Phys. Rev. Lett. {\bf 64} (1990) 2953.
\bibitem{rf:4} Q.~Ouyang and H.~L.~Swinney: Nature {\bf 352} (1991) 610.
\bibitem{rf:5} N.~Maruyama, T.~Koito, J.~Nishida, T.~Sawadaishi, X.~Cieren, K.~Ijiro, O.~Karthaus, and M.~Shimomura: Thin Solid Films {\bf 327} (1998) 854.
\bibitem{rf:6} T.~Tanaka, S.~T.~Sun, Y.~Hirokawa, S.~Katayama, J.~Kucera, Y.~Hirose, and T.~Amiya: Nature {\bf 325} (1987) 796.
\bibitem{rf:7} T.~Hashimoto, M.~Shibayama, and H.~Kawai: Macromolecules {\bf 13} (1980) 1237.
\bibitem{rf:8} H.~Nakanishi, M.~Satoh, and Q.~Tran-Cong-Miyata: Phys. Rev. E {\bf 77} (2008) 020801.
\bibitem{rf:9} M.~Honda, S.~Hozumi, and  S.~Kitayama: Prog. Pacific Polym. Sci. {\bf 3} (1994) 159. 
\bibitem{rf:10} T.~Uematsu, A.~Watanabe, and Y.~Yamaguchi: J. Polym. Sci. B {\bf 42} (2004) 3351.
\bibitem{rf:11} S.~Chatani and T.~Hattori: private communication.
\bibitem{rf:12} M.~Kagami, T.~Yamashita, and H.~Ito: Appl. Phys. Lett. {\bf 79},  (2001) 1079. 
\end{thebibliography}
\end{document}